\documentclass[epj]{webofc}
\usepackage[utf8]{inputenc}
\usepackage[varg]{txfonts}   
\usepackage{booktabs}
\usepackage{xcolor}
\definecolor{darkred}{rgb}{0.4,0.0,0.0}
\definecolor{darkgreen}{rgb}{0.0,0.4,0.0}
\definecolor{darkblue}{rgb}{0.0,0.0,0.4}
\usepackage[bookmarks,linktocpage,colorlinks,
    linkcolor = darkred,
    urlcolor  = darkblue,
    citecolor = darkgreen]{hyperref}
\usepackage{subfigure,slashed,dsfont}
\wocname{EPJ Web of Conferences}
\woctitle{Lattice2017}
\newcommand{\be}{\begin{equation}}
\newcommand{\ee}{\end{equation}}
\newcommand{\M}{\mathcal{M}}

\newcommand{\mpi}{M_{\pi^\pm}}
\newcommand{\fpi}{f_{\pi^\pm}}
\newcommand{\fpip}{f'_{\pi^\pm}}

\begin{document}
%
\selectlanguage{english}
\title{%
Pion decay in magnetic fields
}
\author{%
\firstname{Gunnar~S.} \lastname{Bali}\inst{1,2} 
\and
\firstname{Bastian~B.} \lastname{Brandt}\inst{3}
\and
\firstname{Gergely} \lastname{Endr\H{o}di}
\inst{3}\fnsep\thanks{Speaker, \email{endrodi@th.physik.uni-frankfurt.de}} 
\and
\firstname{Benjamin} \lastname{Gl{\"a}\ss le}\inst{1}
}
\institute{%
Institute for Theoretical Physics, Universit\"at Regensburg, D-93040 Regensburg, Germany
\and
Department of Theoretical Physics, Tata Institute of Fundamental Research, Homi Bhabha Road, Mumbai 400005, India
\and
Institute for Theoretical Physics, Goethe Universit\"at Frankfurt, D-60438 Frankfurt am Main, Germany
}
\abstract{%
The leptonic decay of the charged pion in the presence of background magnetic fields is investigated using quenched Wilson fermions. It is demonstrated that the magnetic field opens up a new channel for this decay. The magnetic field-dependence of the decay constants for both the ordinary and the new channel is determined. Using these inputs from QCD, we calculate the total decay rate perturbatively.
}
\maketitle
\section{Introduction}\label{intro}

Background magnetic fields are known to impact significantly on the physics of 
strongly interacting matter -- affecting the hadron spectrum, creating an anisotropy 
in the ground state and influencing the phase structure of QCD. 
For reviews on this subject and the most recent results on the phase diagram 
see, e.g., Refs.~\cite{Kharzeev:2012ph,Andersen:2014xxa,Endrodi:2015oba}.
Notable examples for systems in nature that exhibit
strong magnetic fields include off-central heavy-ion collisions, the inner core 
of magnetized neutron stars and, possibly, the early stage of the evolution of our universe. 
The magnetic fields in these situations may range up to $B\approx 10^{14-15}\textmd{ T}$, 
so that the interaction between quarks and $B$ becomes as strong as the coupling between 
quarks and gluons. This induces a competition between QCD and QED physics and leads 
to exciting new phenomena.

In the hadronic, confined phase, the response of strongly interacting matter to the magnetic
field $B$ is encoded in the dependence of the 
properties of hadrons, e.g., masses and decay rates, on $B$,
where we assume $\vec{B} = B\,\vec{e}_3$ with $B>0$. 
For the low-energy behavior of the theory, the lightest
hadrons, i.e.\ pions, are of primary interest. The magnetic field couples to electric charge
and thus introduces a splitting between the neutral and charged pions. For the masses $M_{\pi^0}$ and
$M_{\pi^\pm}$,
this splitting
is well known and has been studied in various settings~\cite{Agasian:2001ym,Bali:2011qj,Hidaka:2012mz,Andersen:2012zc,Fayazbakhsh:2012vr,Luschevskaya:2014lga,Avancini:2016fgq,Zhang:2016qrl,Mao:2017wmq,Bali:2017ian}. 
For the neutral pion, the decay rate (i.e.\ the decay constant $f_{\pi^0}$) has also 
been determined using different methods~\cite{Agasian:2001ym,Fayazbakhsh:2012vr,Andersen:2012zc,Fayazbakhsh:2013cha,Avancini:2016fgq,Zhang:2016qrl,Mao:2017wmq}. 

Much less is known about how the magnetic field affects the charged pion decay rate. 
This might be relevant for magnetized neutron stars~\cite{Duncan:1992hi} with pionic degrees of freedom in their core~\cite{Migdal:1990vm}.
In this 
contribution we calculate, for the first time, 
the decay rate using the weak interaction Lagrangian, in the lowest order of 
perturbation theory. 
As we show below, besides the usual decay constant $\fpi$
(which was discussed for $B>0$ using chiral perturbation theory in Ref.~\cite{Andersen:2012zc}), the 
decay rate at $B>0$
involves an additional decay constant that we call $\fpip$. 
This new decay constant has so far been ignored in the literature. 
We perform the perturbative calculation in the so-called lowest Landau-level 
approximation, which is valid for magnetic fields much larger than the squared mass of 
the lepton that is produced in the decay. 
The perturbative treatment 
is then complemented by a non-perturbative determination of both decay constants and of the pion mass.
In particular, we use lattice simulations employing quenched Wilson quarks at $B>0$. Our results 
give direct access to the magnetic field-dependence of the full decay rate
of charged pions into lepton pairs with a muon or an electron.

\section{Pion decay rate and decay constants}\label{sec-1}

CPT invariance ensures that pions with positive and negative electric charge
have the same masses and decay rates. Below we concentrate on the negatively charged pion.
The dominant decay channel is that into a lepton pair
\be
\pi^-(p)\to \ell^-(k) \,\bar\nu_\ell(q)\,,
\label{eq:decay}
\ee
where $\ell^-$ stands for the charged lepton (either electron $\ell=e$ or muon $\ell=\mu$). 
In Eq.~(\ref{eq:decay}) we assigned the momenta $p$, $k$ and $q$ to the pion, the lepton and the antineutrino, 
respectively. The lepton mass will be denoted as $m_\ell$ below. In the effective Lagrangian of the weak interactions, this decay proceeds via 
a four-fermion vertex. The corresponding amplitude is the matrix element of the interaction 
between the initial and final states and reads~\cite{okun2013leptons}
\be
\M = \frac{G}{\sqrt{2}} \cos \theta_c \,L^\mu H_\mu\,,
\label{eq:amplitude}
\ee
where $G$ is Fermi's constant, $\theta_c$ is the Cabibbo angle and $L_\mu$ and $H_\mu$ are the 
matrix elements of the 
leptonic and hadronic contributions to the charged weak current. 
The leptonic factor reads
\be
L^\mu = \bar u_\ell(k)\, \gamma^\mu(1-\gamma^5)\, v_\nu(q)\,,
\label{eq:Lmu}
\ee
where $u_\ell$ and $v_\nu$ are the bispinor solutions of the Dirac equation for the lepton and 
for the antineutrino, respectively. The hadronic factor is defined 
by the matrix element 
\be
H_\mu = \langle0\,|\,\bar d \,\gamma_\mu(1-\gamma^5) u\,|\, \pi^-(p)\big\rangle \,.
\label{eq:Hmudef}
\ee
At zero magnetic field, the parity invariance of QCD dictates that 
the matrix element of the vector part of the weak current vanishes, since it transforms as an 
axial vector (and there is no axial vector in the system). 
Thus, $H_\mu$ is entirely determined by the axial vector part of the weak current and is proportional
to the pion momentum $p_\mu$, being the only Lorentz-vector in the problem. The proportionality 
factor defines the pion decay constant $\fpi$. 
Note that there exist subtleties regarding the definition of
decay constants of electrically charged hadrons, see, e.g.,
Ref.~\cite{Patella:2017fgk}. This however goes
beyond the tree-level considerations of this presentation.

In the presence of a background magnetic field
(in general a Lorentz-tensor $F_{\mu\nu}$), 
further Lorentz structures that are relevant
for weak V-A decays can be formed: the vector $F_{\mu\nu}p^\nu$ and 
the axial vector $-i\epsilon_{\mu\nu\rho\sigma}F^{\nu\rho}p^\sigma/2$. Specifically, the 
matrix element of the vector part of the weak current can be nonzero. 
For a pion at rest, $p^\mu=(p_0,\vec{p})=(\mpi,0)$, and a magnetic field oriented 
in the positive $z$ direction, $F_{12}=-F_{21}=B$, the matrix element 
becomes
\be
H_\mu = e^{i\mpi t}\left[ \fpi \mpi \delta_{\mu0} +i \fpip eB \mpi \delta_{\mu3} \right]\,,
\label{eq:Hmu}
\ee
where we adopt the
normalization convention $\fpi\approx 130 \textmd{ MeV}$ in the vacuum with physical quark masses, and 
we measure the magnetic field (which has mass dimension two) in units of the elementary electric charge $e>0$. 
Note that while the complex phase of the individual terms in~(\ref{eq:Hmu}) is a matter of convention, their ratio 
is physical and follows from the transformation properties of~(\ref{eq:Hmudef}) under discrete symmetries. In our convention both decay constants are real.

Using the amplitude~(\ref{eq:amplitude}), the decay rate is given by Fermi's golden rule
\be
\Gamma = \int\!\textmd{d}\Phi \sum_{\{s\}} |\M\,|^2\,,
\ee
involving an integral over the phase space $\Phi$ and a sum over the intrinsic quantum numbers $s$
of the outgoing particles. 
For $B=0$ the latter involve the spins $s_\ell$ and $s_\nu$ of the leptons. 
To carry out the sum over these, we need the spin sums for the bispinors in~(\ref{eq:Lmu}). For the charged lepton, 
this reads
\be
\sum_{s_\ell} u_{\ell}^{s_\ell}(k) \,\bar u_{\ell}^{s_\ell}(k) = \slashed{k}+m_\ell\,, 
\label{eq:spinsum0}
\ee
while for the neutrino the sum gives $\slashed{q}$, neglecting the tiny neutrino masses.
A textbook calculation~\cite{okun2013leptons} finally results in
\be
\Gamma(B=0) = \frac{(G\cos\theta_c)^2}{8\pi}\,|\fpi(0)\,|^2\, [\mpi^2(0)-m_\ell^2]^2\frac{m_\ell^2}{\mpi^3(0)}\,,
\ee
where we indicated that $\fpi$ and $\mpi$ are understood at $B=0$.

\subsection{Decay rate for \boldmath$B>0$}

For $B>0$ the above calculation 
is modified considerably. Here we do not describe the details, which will be included in a forthcoming
publication~\cite{to_come}, but merely summarize the prime points of the calculation.
First of all, we need to include 
the second contribution to $H_\mu$ from Eq.~(\ref{eq:Hmu}). In addition, the 
bispinor solution $u_\ell$ is affected, since the Dirac equation for the charged lepton 
depends on the magnetic field through the electromagnetic vector potential. 
In particular, the solutions are quantized and correspond to the so-called Landau levels~\cite{landau1977quantum} -- states with 
definite angular momentum in the $z$ direction, labeled 
by the Landau index $n\in\mathds{Z}^+_0$. 

Each Landau level carries a degeneracy 
proportional to the flux $eB\cdot L^2$ of the magnetic field through the area $L^2$ 
of the system. 
Moreover, the energies of the states corresponding to the $n$-th Landau level are bounded from below by 
$\sqrt{m_\ell^2+2n eB}$. 
This allows us to simplify the problem by considering the limit $eB\gg m_\ell^2$. 
For such strong magnetic fields, 
only the states with $n=0$ will contribute, enabling us to restrict
ourselves to leptons in the lowest Landau level (LLL) and to neglect
levels with $n>0$.
The LLL states\footnote{For the role of LLL states in QCD with $B>0$, see Ref.~\cite{Bruckmann:2017pft}.} are special because they are effectively one-dimensional, allowing 
only motion aligned with the magnetic field, and also because the spin $s_\ell$ is fixed to be antiparallel to the magnetic field for them.

Taking this into account, the equivalent of Eq.~(\ref{eq:spinsum0}) reads
\be
\sum_{\rm LLL} u_{\ell}^{s_\ell}(k) \,\bar u_{\ell}^{s_\ell}(k) = (\slashed{k}_\parallel+m_\ell) \cdot 
\frac{1-\sigma^{12}}{2} \cdot \frac{eB\cdot L^2}{2\pi}\,.
\label{eq:spinsumB}
\ee
Here, $\slashed{k}_\parallel=k^0\gamma^0-k^3\gamma^3$ emerges due to the one-dimensional nature
of the LLL states, and the second factor involving the relativistic spin operator $\sigma^{12}=i\gamma^1\gamma^2$ projects to states with negative spin. Finally, the third 
factor results from the sum over the LLL-degeneracy. 
Note that the corresponding spin sum for the neutrino is not affected by the magnetic field. 
Using the spin sum~(\ref{eq:spinsumB}), we obtain for the decay rate,
\be
\Gamma(B) 
= 
eB\,\frac{(G\cos\theta_c)^2}{2\pi}\,|\fpi(B)+i\fpip(B) eB\,|^2\frac{m_\ell^2}{\mpi(B)}\,.
\label{eq:resB}
\ee
Note that we can assume
$G\cos\theta_c$ to be independent of $B$, since in this context
the latter is a low energy scale, being much smaller than the squared mass of
the $W$ boson that mediates the weak interaction.

A sensible way to quantify the impact of the magnetic field
on $\Gamma$ is to look at the ratio
\be
\frac{\Gamma(B)}{\Gamma(0)} = 4 \frac{|\fpi(B)\,|^2+|\fpip(B)\,eB\,|^2}{|\fpi(0)\,|^2} \\
\cdot \left(1-\frac{m_\ell^2}{\mpi^2(0)}\right)^{-2} \!\frac{eB}{\mpi^2(0)} \,\frac{\mpi(0)}{\mpi(B)}\,,
\label{eq:Gammarat}
\ee
in which the constants $G$ and $\theta_c$ cancel.
Remember that this result is only valid 
in the limit $eB\gg m_\ell^2$.

\section{Lattice simulations}\label{sec-2}

To calculate the ratio $\Gamma(B)/\Gamma(0)$ of Eq.~(\ref{eq:Gammarat}), it remains to determine 
the non-perturbative parameters $\fpi(B)$, $\fpip(B)$ and $\mpi(B)$ in QCD. 
To this end we perform lattice simulations using quenched Wilson quarks. 
We work with the zero-temperature 
lattice ensembles generated in Ref.~\cite{Bali:2017ian}, with parameters listed in Tab.~\ref{tab-1}.

\begin{table}[thb]
\small
\centering
\caption{The parameters of our lattice ensembles: lattice size, inverse gauge coupling $\beta$ and 
lattice spacing $a$.}
\label{tab-1}
\begin{tabular}{c|c|c}\toprule
 $N_s^3\times N_t$ & $\beta$ & $a$ \\\midrule
 $12^3\times 36$ & $5.845$ & $0.124\textmd{ fm}$ \\
 $16^3\times 48$ & $6.000$ & $0.093\textmd{ fm}$ \\
 $24^3\times 72$ & $6.260$ & $0.062\textmd{ fm}$ \\\bottomrule
\end{tabular}
\end{table}

For this first study, we performed measurements with various different valence quark masses 
so that the $B=0$ pion mass spans the range $415\textmd{ MeV}\le \mpi \le 770\textmd{ MeV}$. The magnetic 
field is varied in the window $0\le eB<1.5 \textmd{ GeV}^2$. 
Note that the magnetic field modifies the critical bare mass parameter for Wilson fermions 
by large lattice artefacts, as was first pointed out in Ref.~\cite{Bali:2015vua}. We correct for this by tuning the bare mass parameter along 
the magnetic field-dependent line of constant physics determined in Ref.~\cite{Bali:2017ian}.

\begin{figure}[t]
  \centering
  \includegraphics[width=8cm,clip]{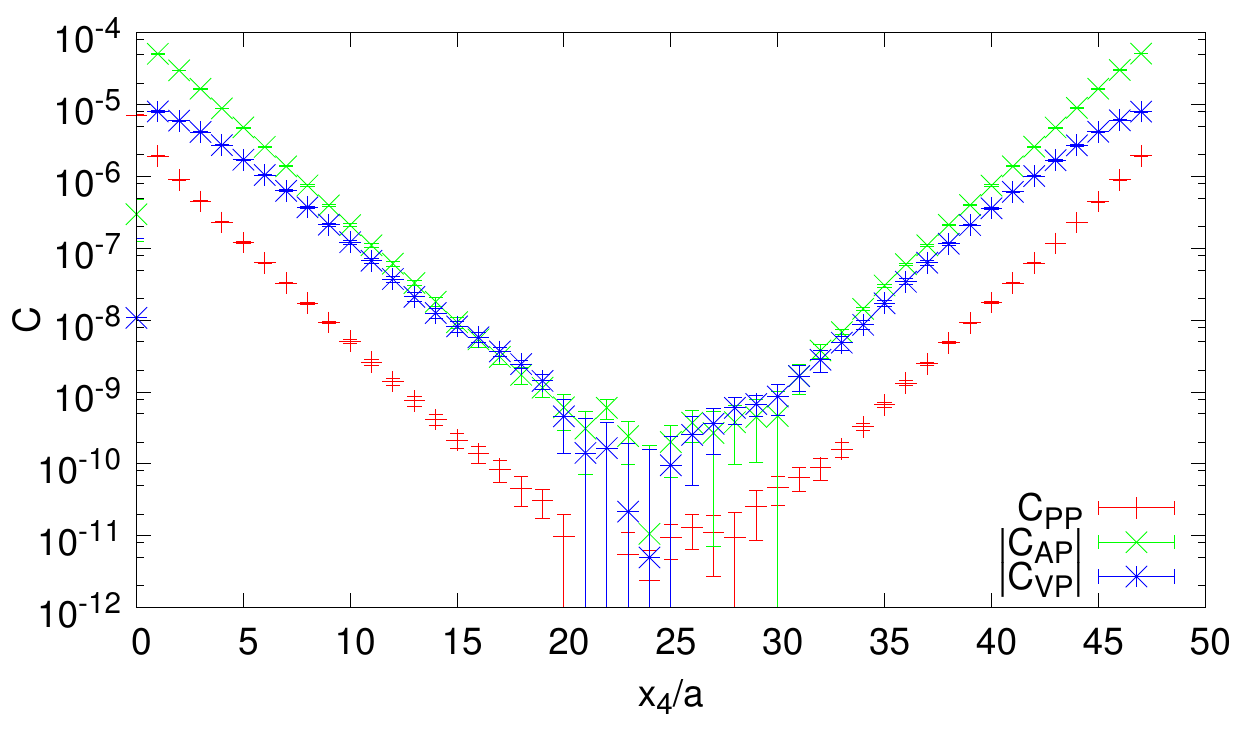}
  \caption{Pseudoscalar-pseudoscalar (red), pseudoscalar-axial vector (green) and pseudoscalar-vector
  (blue) correlators, measured on our $16^3\times48$ ensemble. Note that $C_{AP}$ and $C_{VP}$ 
  are odd under time reversal and change
  sign at $x_4/a=N_t/2$; the plot shows the modulus of these correlators. }
  \label{fig:corr}
\end{figure}

To calculate the pion mass and the decay constants we investigate the matrix elements 
$H_0$ and $H_3$, defined in Eq.~(\ref{eq:Hmudef}). To this end, we create a pion state using the  
smeared pseudoscalar operator $P^{\dagger}=\bar d \gamma^5 u$ at the source, and annihilate it at the sink 
using either $P$, the local axial vector operator
$A=\bar u \gamma_0 \gamma^5 d$ or the local vector operator $V=\bar u \gamma_3 d$, to obtain 
the correlators $C_{PP}$, $C_{AP}$ or $C_{VP}$, respectively. 
For more details
about our measurement strategy, see Ref.~\cite{Bali:2017ian}. 
It follows from Eq.~(\ref{eq:Hmu}) that in Euclidean space-time all three correlators 
decay exponentially as $\exp(-\mpi x_4)$, where $x_4>0$.
This is demonstrated in Fig.~\ref{fig:corr} for $eB=1.3\textmd{ GeV}^2$
and a zero-magnetic field pion mass of $\mpi=770\textmd{ MeV}$. 

We fit the three correlators simultaneously (for sufficiently large
values of $x_4$) using the functions
\be
C_{\mathcal{O} P}(t) = c_{\mathcal{O}P}
\left[ e^{-\mpi x_4} \pm e^{-\mpi(N_ta-x_4)} \right]\,, \quad\quad \mathcal{O}=P,A,V\,,
\ee
where the positive sign in front of the second term is taken for $\mathcal{O}=P$ and the negative sign
for $\mathcal{O}=A$ and $\mathcal{O}=V$.
The decay constants are obtained from the amplitudes of the exponential decays as
\be
\fpi = Z_A\cdot \frac{\sqrt{2} \,c_{AP}}{\sqrt{\mpi c_{PP}}}, \quad\quad
i\fpip eB = Z_V\cdot \frac{\sqrt{2}\, c_{VP}}{\sqrt{\mpi c_{PP}}}\,,
\ee
where $Z_A$ and $Z_V$ are the multiplicative renormalization constants of the axial vector and vector 
currents, respectively. Up to lattice artefacts, these ultraviolet quantities are expected to be independent of the 
magnetic field (which is an infrared parameter). We employ the $B=0$ non-perturbative results of Ref.~\cite{Gimenez:1998ue} 
(see also Ref~\cite{Gockeler:1998ye}) and fit these in combination with the asymptotic perturbative two-loop results of~\cite{Skouroupathis:2008mf}
(see also Ref~\cite{Bali:2013kia}) to a Pad\'e parametrization.

\begin{figure}[ht!]
 \centering
 \includegraphics[width=7cm]{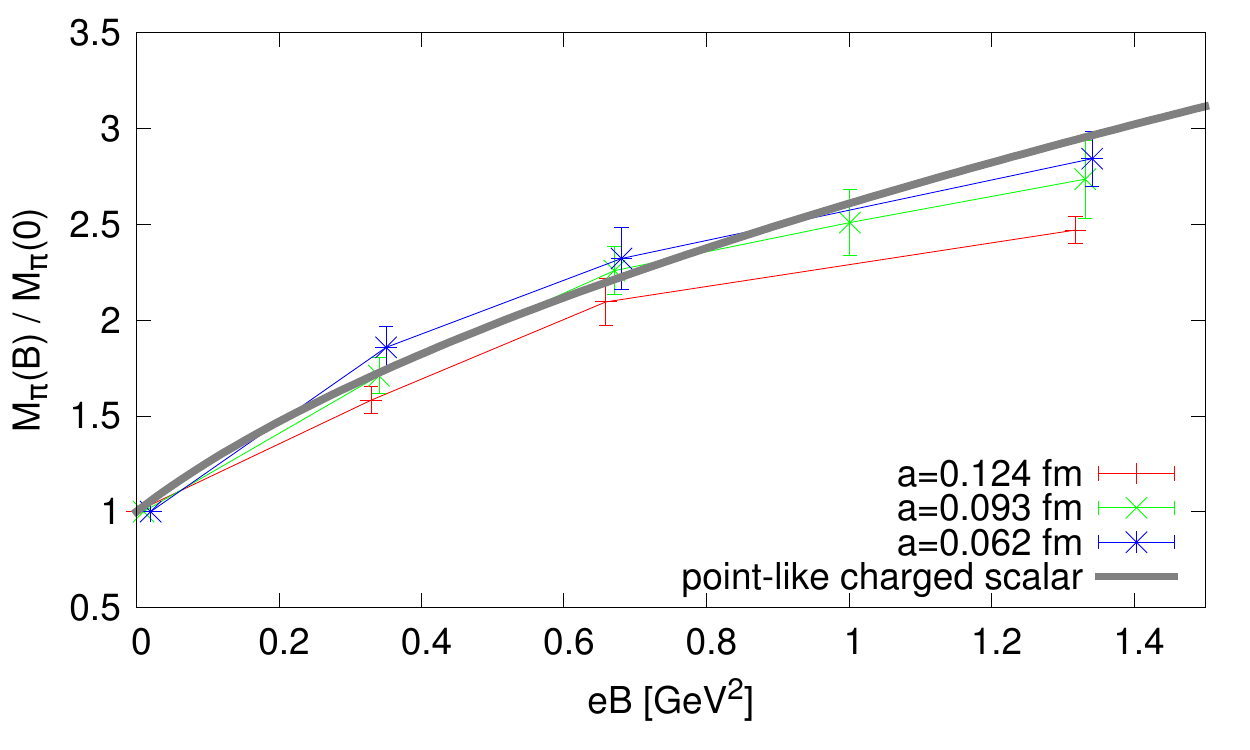}\;
 \includegraphics[width=7cm]{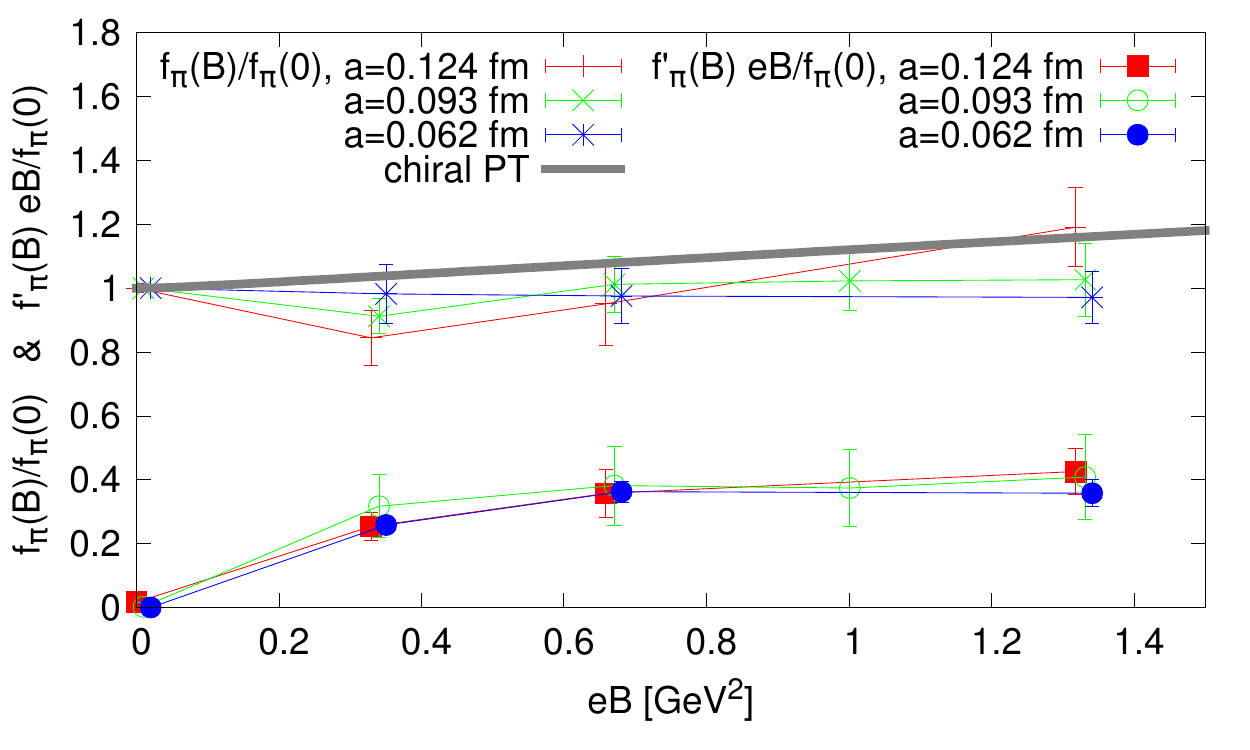}
 \caption{\label{fig:mf}The mass of the charged pion (left panel) 
 and the two decay constants $\fpi$ and $\fpip eB$ (right panel) as functions of the 
 magnetic field for a $B=0$ pion mass of $415\textmd{ MeV}$, as measured on 
 three different lattice ensembles. The solid curve in the left panel depicts the mass 
 of a point-like charged scalar particle that interacts only with the magnetic field. 
 The solid curve in the right panel shows the prediction of chiral perturbation theory for $\fpi(B)$~\cite{Andersen:2012zc}. The lines connecting the data points merely serve 
 to guide the eye. The points have been shifted horizontally for better visibility.}
\end{figure}

The results for the pion mass and for the decay constants -- in units of $\mpi(0)$ and 
of $\fpi(0)$, respectively -- are shown in Fig.~\ref{fig:mf} for a $B=0$ pion mass
of $415 \textmd{ MeV}$. 
For $\mpi$, the three lattice spacings lie reasonably close to each other, indicating small 
discretization errors, at least for $eB<0.8\textmd{ GeV}^{2}$. (Note that without the magnetic field-dependent tuning 
of the bare quark mass~\cite{Bali:2017ian}, lattice artefacts
would be larger.)
For comparison, we also consider the mass of a point-like scalar
particle of charge $e$, $M_{\phi}(B)=\sqrt{\mpi^2(0)+eB}$, which lies
quite close to our results for $\mpi(B)$. This trend has already been observed in the literature,
both using dynamical staggered~\cite{Bali:2011qj}, quenched Wilson~\cite{Hidaka:2012mz} and quenched overlap quarks~\cite{Luschevskaya:2014lga}.

Our results for the decay constants are plotted in the right panel of 
Fig.~\ref{fig:mf}. $\fpi$ shows no significant dependence on $B$ 
for the range of magnetic fields that we study here. We also have a clear signal for $\fpip eB$, 
showing an initial linear increase, followed by a plateau as $B$ 
grows. The behavior at low magnetic fields is described by a linear coefficient 
$\fpip/\fpi\approx 0.8\textmd{ GeV}^{-2}$. 
In the right panel of Fig.~\ref{fig:mf} we also include the prediction for $\fpi$ from 
chiral perturbation theory~\cite{Andersen:2012zc}, which suggests a gradual enhancement of $\fpi(B)$ and is consistent with our results within our present uncertainties of about $10\%$.

\begin{figure}[ht!]
 \centering
 \includegraphics[width=8cm]{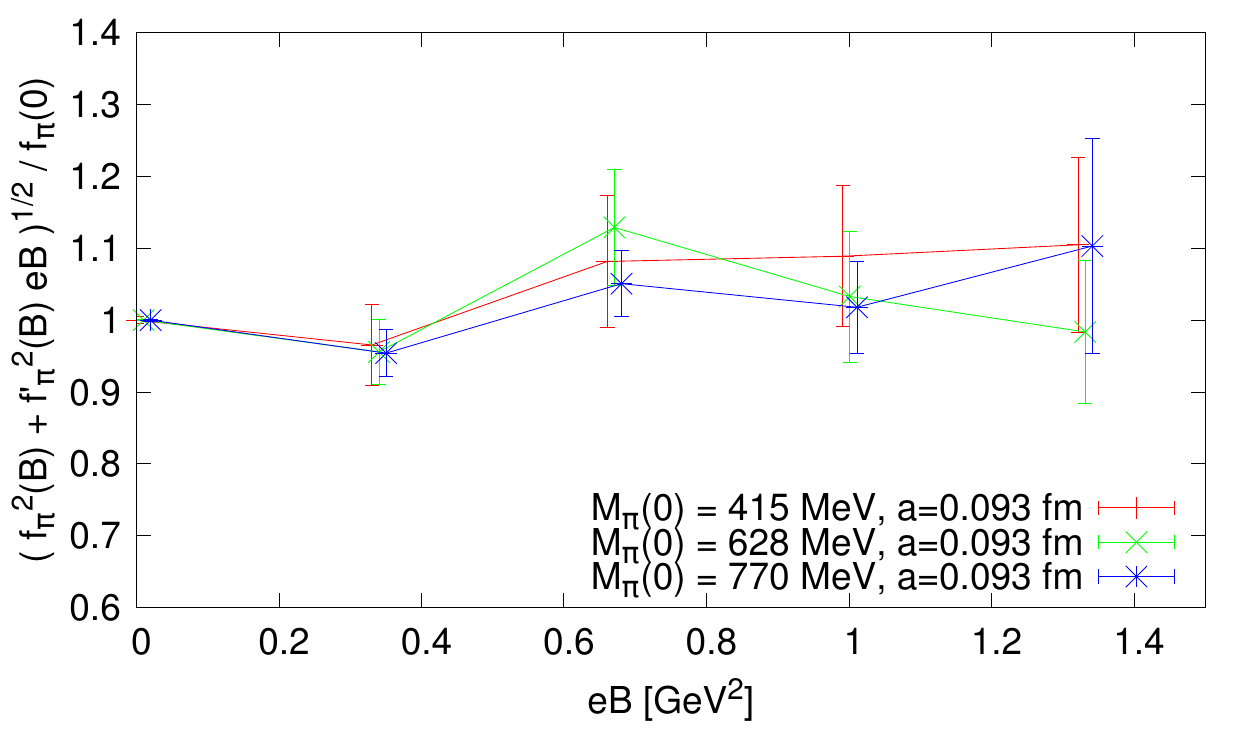}
 \caption{\label{fig:Gr}The magnetic field-dependence of the 
 combination $\sqrt{|\fpi(B)|^2+|\fpip(B)eB|^2}/\fpi(0)$, which appears in
 the decay rate~\protect(\ref{eq:Gammarat}) for different $B=0$ pion masses. 
 The points have been shifted horizontally for better visibility.}
\end{figure}

Next we investigate the dependence of the results on the quark mass. We consider three 
sets of measurements with $B=0$ pion masses of $415 \textmd{ MeV}$, $628\textmd{ MeV}$ and $770 \textmd{ MeV}$. In Fig.~\ref{fig:Gr} we plot the combination $\sqrt{|\fpi(B)|^2+|\fpip(B) \,eB|^2}/\fpi(0)$ -- 
the square of which characterizes the magnetic field-dependence of the full decay rate, Eq.~(\ref{eq:Gammarat}). 
The plot reveals that to our current accuracy, quark mass effects for our $a=0.093\textmd{ fm}$ ensemble are invisible in this particular combination. Our results for the other 
lattice spacings show the same behavior.

\begin{figure}[t]
 \centering
 \includegraphics[width=7cm]{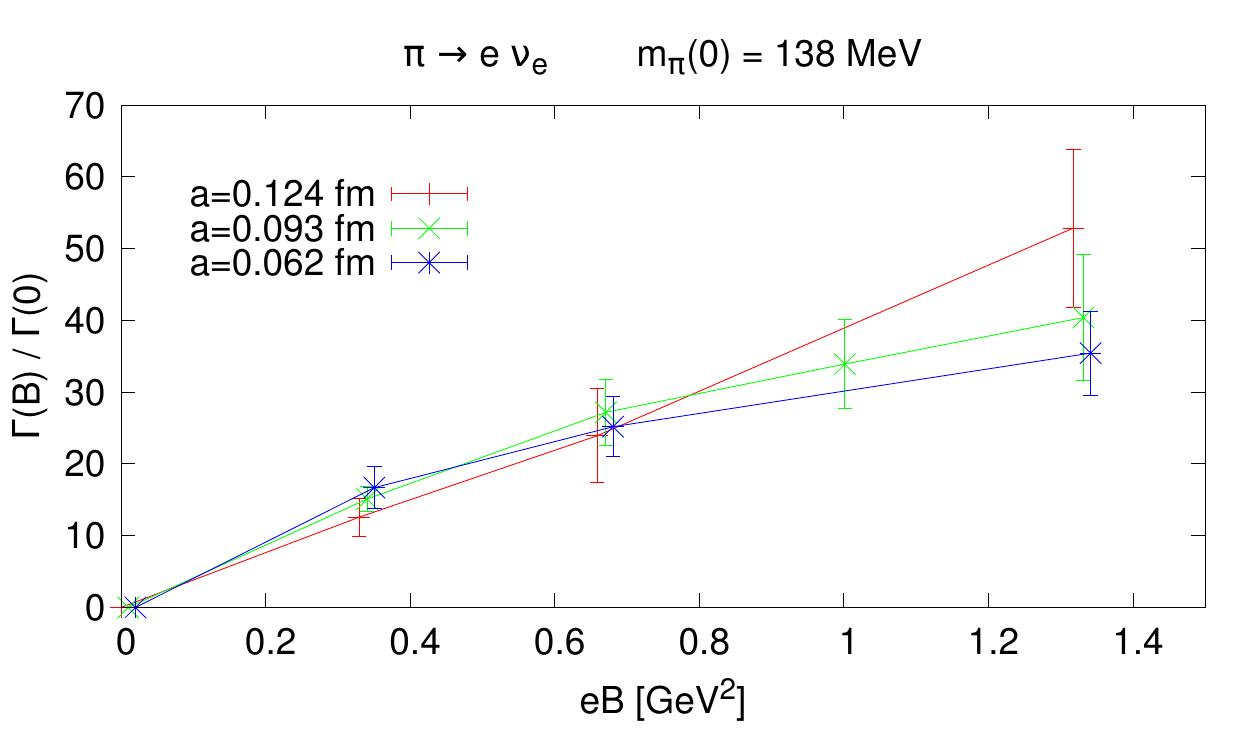}\,
 \includegraphics[width=7cm]{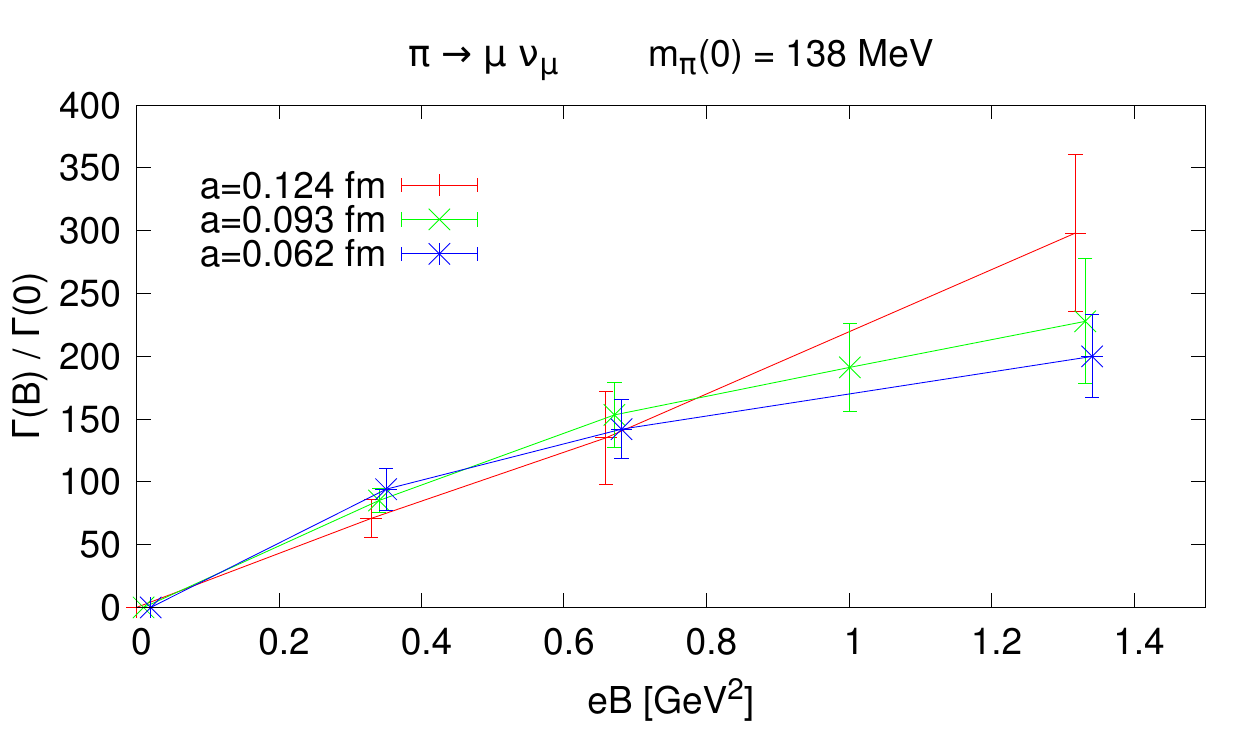}
 \caption{\label{fig:dr}The decay rate of the process $\pi^-\to e^-\bar\nu_e$ (left panel) 
 and for $\pi^-\to\mu^-\bar\nu_\mu$ (right panel) in units of the respective $B=0$ values. 
 Here we took 
 the physical pion mass in the formula~\protect(\ref{eq:Gammarat}) and assumed -- 
 as Fig.~\ref{fig:Gr} suggests -- that 
 the QCD factor does not change considerably between the physical point and $\mpi=415 \textmd{ MeV}$. The points have been shifted horizontally for better visibility.}
\end{figure}

Finally, we speculate about the decay rate at the physical point, $\mpi(0)=138 \textmd{ MeV}$. 
Assuming that the QCD factor is not significantly modified as the quark mass is reduced further, 
we take our results for the lowest quark mass and insert 
these into the formula for the full decay rate~(\ref{eq:Gammarat}), however, replacing the pion mass with $\mpi(0)=138\textmd{ MeV}$. We use the free-case formula 
$\mpi(B)=\sqrt{\mpi^2(0)+eB}$, which describes our results sufficiently accurately, as we 
demonstrated in the left panel of Fig.~\ref{fig:mf}. The so-obtained estimate for the decay rate 
of the charged pion at the physical point is shown in Fig.~\ref{fig:dr}. In the left panel we consider the decay 
into electrons $\ell=e$, whereas the right panel depicts the results for the decay into muons $\ell=\mu$.
The only difference between the two cases is the lepton mass $m_e=0.5 \textmd{ MeV}$ versus 
$m_\mu=105\textmd{ MeV}$. Remember that since we employed the lowest Landau-level 
approximation, the decay rate formula is only valid for magnetic 
fields well above the respective squared lepton masses, which is the case for
all our data points.

The results reveal a fast approach towards the continuum limit. 
For magnetic fields of the order $eB\approx0.3 \textmd{ GeV}^2$ -- which is a typical value 
relevant for the physical situations discussed in the introduction --
a drastic enhancement of $\Gamma$ is clearly visible. For the electronic decay the rate 
enhances by a factor $\sim20$ with respect to $\Gamma(0)$, 
while for the muonic decay the enhancement is by more than two orders 
of magnitude. Thus, the average lifetime of the charged pion is reduced drastically.

We point out that the
ratio of the electronic and the muonic decay rates is independent of the magnetic field,
\be
eB\gg m_\mu^2:\quad\quad \frac{\Gamma(\pi\to e \bar\nu_e)}{\Gamma(\pi\to \mu \bar\nu_\mu)} 
= \frac{m_e^2}{m_\mu^2} \approx 2.27\cdot 10^{-5}\,,
\label{eq:GRAT}
\ee
and is suppressed by a factor of five compared to the $B=0$ ratio~\cite{Agashe:2014kda}. 
Therefore, for strong background magnetic fields the muonic decay becomes even more dominant
than it already is at $B=0$. 

\section{Conclusions}

In this contribution we studied the mass and the decay rate of charged pions in the 
presence of background magnetic fields. To determine the rate of decay 
into leptons, we have performed a 
tree-level perturbative calculation in the electroweak theory, assuming that the magnetic field
is well above the squared mass of the charged lepton, $eB\gg m_\ell^2$. This assumption 
enables the use of the lowest Landau-level approximation for the outgoing lepton state, which 
simplifies the calculation considerably. 

The perturbative calculation involves the matrix element of the weak current between the vacuum 
and a pion state.
We have demonstrated for the first time, that 
for nonzero background fields this matrix element is characterized 
by two independent decay constants $\fpi$ and $\fpip$. 
Both decay constants were determined non-perturbatively on the lattice using quenched Wilson 
fermions. We investigated the dependence of the results on the lattice spacing and on 
the quark mass (or, equivalently, on the $B=0$ pion mass), revealing modest discretization errors
and quark-mass effects.

We find the muonic partial decay rate, which amounts to over 99.9\% of
the total decay rate, to be enhanced by two orders of magnitude for
magnetic fields of around $eB\approx 0.3\textmd { GeV}^2$, that are of phenomenological
relevance. We predict a similar enhancement by a factor of around 20
for the electronic decay rate. The ratio~(\ref{eq:GRAT}) of the decay rates in the two channels is found to be independent of $B$, and of 
all non-perturbative quantities -- thus this prediction applies irrespective of the results of our lattice simulations.
Our findings may have applications for the physics of magnetars.
\\

\noindent
{\bf Acknowledgments}
This research was funded by the DFG (Emmy Noether Programme EN 1064/2-1 and
SFB/TRR 55). BB acknowledges support from
the Frankfurter F{\"o}rderverein f{\"u}r Physikalische Grundlagenforschung.
We thank Zolt\'an Fodor and Andreas Sch{\"a}fer for illuminating 
discussions.

\bibliography{lattice2017}

\end{document}